\documentclass[12pt]{iopart}
\usepackage{iopams}  
\usepackage[dvipdfmx]{graphicx}
\begin{document}

\title[Transfer dynamics of quantum teleportation]{An investigation of the transfer dynamics of quantum teleportation by weak measurement statistics}

\author{Masanori Hiroishi$^1$ and Holger F.~Hofmann$^{1,2}$}

\address{$^{1}$Graduate School of Advanced Sciences of Matter, Hiroshima University, Kagamiyama 1-3-1, Higashi Hiroshima 739-8530, Japan \\
$^{2}$JST, CREST, Sanbancho 5, Chiyoda-ku, Tokyo 102-0075, Japan}
\ead{hofmann@hiroshima-u.ac.jp}
\begin{abstract}
We explore the mechanism of quantum teleportation by analyzing the weak measurement statistics post-selected by the result of the Bell measurement for the joint system composed of the input $A$ and the spatially separated output $B$. It is shown that the weak measurement statistics observed before the Bell measurement includes correlations which relate every physical property in the input $A$ to a corresponding physical property in the output $B$. The Bell measurement thus identifies the accidental relation between $A$ and $B$ already present in the quantum fluctuations of the input state. Significantly, this relation applies to all physical properties equally, and is completely independent of the input state. Teleportation therefore copies all physical properties of input system $A$ to output system $B$, irrespective of whether the input state is an eigenstate of the property or not.
\end{abstract}

\pacs{
03.65.Ta, 
03.67.-a, 
}

\maketitle

\section{Introduction}

Quantum teleportation is one of the fundamental applications of entanglement in quantum information \cite{Ben93}. In the teleportation process, a maximally entangled state of two systems, $R$ and $B$, is distributed to the sender and the receiver. The sender then combines the input system $A$ with system $R$ of the entangled pair to perform a Bell measurement, which is described by a projection on a maximally entangled state of $A$ and $R$. As a result of this projection, the unknown quantum state of $A$ is transferred to the remote system $B$, which now carries quantum coherences that correspond precisely to the quantum coherences of $A$ before the Bell measurement. 

Although the Hilbert space formalism of the teleportation process is mathematically simple and consistent, it describes non-local effects of measurements on an entangled state. The origin of these effects is not obvious, since the quantum measurement process combines the physical disturbance of properties by interaction with changes to the statistics caused by additional information gained in the process. It may therefore be possible to explain the non-local effects of teleportation entirely in terms of subjective information gain, as indicated by the description of continuous variable quantum teleportation using Wigner functions \cite{Bra98}. In that case, the analogy between Wigner functions and classical phase space distributions can be used to construct a classical model of teleportation, where the Bell measurement merely uncovers the correct value of the phase space distance that describes the relation between the physical properties of input system $A$ and output system $B$ before the Bell measurement. The projection of the state vector thus appears as a process that modifies the information about system $B$ without altering any of its physical properties. This viewpoint may also apply to the general quantum teleportation of $d$-level systems, since there is no experimentally accessible evidence of a change of the physical properties in $B$ caused by the performance of the Bell measurement in $A$ \cite{Hof02}.

The standard formalism of quantum mechanics does not provide any direct answer to the question of how and when the physical properties are transferred from system $A$ to system $B$ in a quantum teleportation. It has even been suggested that teleportation corresponds to time travel, where the system $R$ is the channel that transfers the state back in time, so that it is available in $B$ at the moment of entanglement generation \cite{Llo11a,Llo11b}. Although this seems to be a rather extreme interpretation of the formalism, it is based on the fundamental observation that the teleportation process relies on a time-symmetric combination of entangled state preparation and Bell state measurement. The conventional formulation of quantum mechanics tends to hide this symmetry by overemphasizing the role of the initial state, while neglecting the implications of measurement results for the situation before the measurement. To correct this inadvertent bias, it may be useful to apply a time-symmetric description to the situation between state preparation and Bell measurement. Such a time-symmetric description is provided by the post-selected measurement statistics obtained in weak measurement, which give equal importance to the initial and the final state of a quantum process \cite{Aha88}. 
 
It has already been shown that the statistics obtained by weak measurements can be used to identify the disturbance of a physical system in a general measurement by separating the actual changes of a physical property from the effects of new information about these properties \cite{Mir07,Lun10,Hof10}. In the same spirit, weak measurements could be applied to identify the physical properties of the systems involved in the teleportation before the Bell measurement is performed. In the following, we will analyze the complete statistics of the systems $A$, $R$, and $B$ between state preparation and Bell measurement, taking into account the post-selection of a specific result of the Bell measurement. In this analysis, the effects of the Bell measurement correspond to a Bayesian update of the initial state that selects the appropriate sub-ensemble from the total statistics without any physical changes to the properties of the systems. The weak measurement analysis of quantum teleportation thus shows that the correlations between the physical properties of input $A$ and output $B$ that enable teleportation can already be found in the quantum fluctuations of the initial state before the Bell measurement is actually performed.

The rest of the paper is organized as follows. In section \ref{sec:tele}, a general formulation of the teleportation process is given and the observable effects of the Bell measurement are considered. In section \ref{sec:weak}, the statistics of weak measurements is introduced. In section \ref{sec:local}, it is shown how the input state determines the statistics of local weak measurements on any one of the three systems in the interval between entanglement preparation and Bell measurement. In section \ref{sec:precorr}, the correlations between input and output are analyzed. The results show that all physical properties in $A$ and in $B$ are perfectly correlated, as indicated by a weak value of zero for projections onto mutually orthogonal states in $A$ and in $B$. In section \ref{sec:partial}, it is shown that partial teleportation by a quantum controlled Bell measurement leaves the correlations between $A$, $R$, and $B$ visible in the final output, where the ideal correlations of teleportation are replaced by the correlations of optimal cloning. In section \ref{sec:discuss}, the implications of the results for the teleportation process are discussed and a statistical interpretation is considered. Section \ref{sec:conclude} summarizes the results and concludes the paper. 

\section{Elements of quantum teleportation}
\label{sec:tele}

\begin{figure}[th]
\centering
\vspace{-4cm}
 \includegraphics[width=125mm]{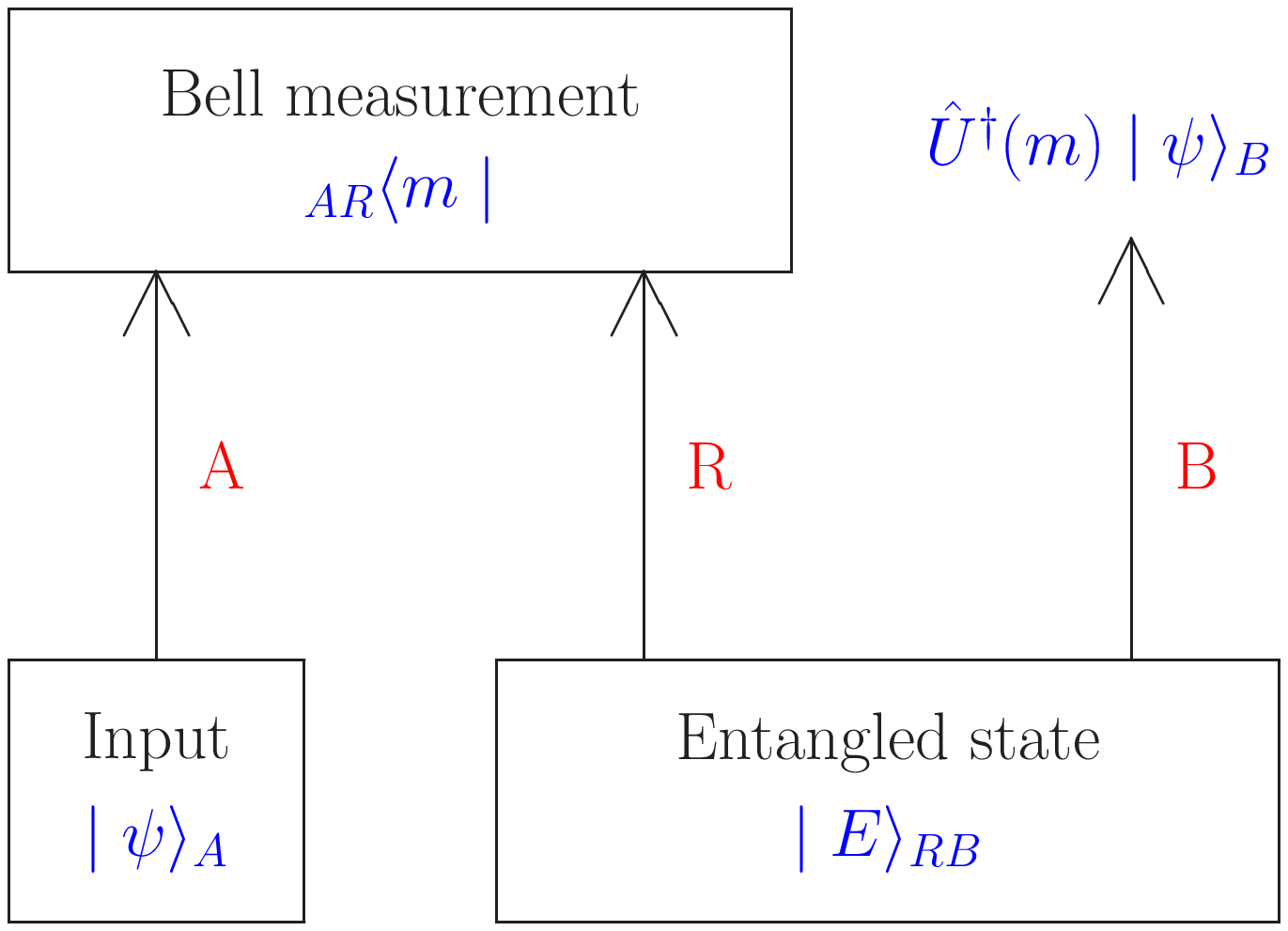}
\vspace{-6.5cm}
   \caption{Illustration of quantum teleportation.The outcome of the Bell measurement defines the unitary transform $\hat{U}(m)$ that relates the output state in $B$ to the unknown input in $A$.}
   \label{fig1}
\end{figure}

We consider the teleportation of the arbitrary quantum state $\mid \psi \rangle$ of a $d$-level system, as illustrated in Fig. \ref{fig1}. Initially, system $A$ carries the input state, and systems $R$ and $B$ are in a maximally entangled state,
\begin{equation}
\label{eq:E}
\mid E \rangle_{RB} = \frac{1}{\sqrt{d}} \sum_n \mid n; n \rangle_{RB}.
\end{equation} 
The quantum information of the initial state $\mid \psi \rangle$ can then be transfered to the remote system $B$ by a joint measurement of systems $A$ and $R$. This measurement is described by a projection onto an orthogonal basis of maximally entangled states, also known as a Bell basis. Since maximally entangled states can be transformed into each other by local unitary transformations, it is possible to express the $d^2$ states $\mid m \rangle$ of the Bell basis in terms of the initial entangled state $\mid E \rangle$ by
\begin{equation}
\label{eq:Bellstate}
\mid m \rangle_{AR} = \hat{U}(m) \otimes I \; \mid E \rangle_{AR},
\end{equation}
where the $d^2$ unitary operations $\hat{U}(m)$ are orthogonal in terms of their product traces, 
\begin{equation}
\mbox{Tr} \left(\hat{U}^\dagger (m) \hat{U}(m^\prime) \right)=\delta_{m,m^\prime}.
\end{equation}
The Bell measurement results in the projection of the initial state in $A$, $R$, and $B$ described by the partial inner product in $AR$,
\begin{eqnarray}
\label{eq:tele}
_{AR} \langle m \mid \psi; E \rangle_{A;RB} &=& \frac{1}{d} \sum_n \langle n \mid \hat{U}^\dagger(m) \mid \psi \rangle \mid n \rangle_B
\nonumber \\
&=& \frac{1}{d} \; \hat{U}^\dagger(m) \mid \psi \rangle_B.
\end{eqnarray}
After the Bell measurement, the state of system $B$ is given by a well-defined unitary transformation of the initial state $\mid \psi \rangle$. The effect of this unitary can be compensated by applying the unitary $\hat{U}(m)$ that describes the transformation of the entangled state $\mid E \rangle$ into the Bell state $\mid m \rangle$. 

Although the measurement formalism seems to describe an instantaneous change of the state, the Bell measurement performed at $A$ does not change the total statistics observable in system $B$. For random outcomes of the Bell measurement, the average quantum state in $B$ is
\begin{equation}
\label{eq:decomp}
\sum_m \frac{1}{d^2} \hat{U}^\dagger(m) \mid \psi \rangle \langle \psi \mid \hat{U}(m)
= \frac{1}{d} \hat{I},
\end{equation}  
which is equal to the result obtained by tracing out $A$ and $R$ before the measurement. It is therefore possible to describe the local state in $B$ before the Bell measurement as a mixture of the states conditioned by the different outcomes of the Bell measurement \cite{Hof02}. However, it is important to note that the decomposition in Eq.(\ref{eq:decomp}) works for any state $\mid \psi \rangle$, so it is still necessary to explain how the Bell measurement can fix the choice of a particular state $\mid \psi \rangle$. For this purpose, it is useful to consider the correlations between $A$ and $B$ in more detail.

\section{Weak measurements and the transient state}
\label{sec:weak}

A direct observation of correlations between the input $A$ and the output $B$ before the Bell measurement is impossible because a measurement of $A$ will alter the state and therefore change the results of the Bell measurement. The teleported state would simply be the output state of the measurement performed in $A$, and no insights into the actual teleportation process would be gained. However, it is possible to perform a weak measurement, where the measurement interaction is so small that the changes to the quantum state can be neglected. The results of these measurements are equally determined by initial state $\mid i \rangle$ and post-selected final measurement projection $\mid f \rangle$ and can be expressed in terms of weak values \cite{Aha88}. For an observable described by the operator $\hat{V}$, the weak value is 
\begin{equation}
\label{eq:wv}
\frac{\langle f \mid \hat{V} \mid i \rangle}{\langle f \mid i \rangle} = \mbox{Tr}\left(\hat{V} \; \frac{\hat{\rho}_i \hat{\Pi}_f}{\mbox{Tr}\left(\hat{\rho}_i \hat{\Pi}_f \right)}\right)
\end{equation}
The right hand side of this equation shows the general expression of weak values using the density operator $\hat{\rho}_i$ as initial state and the measurement operator $\hat{\Pi}_f$ for the final measurement. In this formulation, it is easy to see that the weak values can be interpreted as expectation values of a transient state \cite{Hof10,Shi10}. For complex weak values, this transient state is given by the ordered product of density operator and measurement operator, normalized to a trace of one,
\begin{equation}
\hat{T}(f)=\frac{\hat{\rho}_i \hat{\Pi}_f}{\mbox{Tr}\left(\hat{\rho}_i \hat{\Pi}_f \right)}.
\end{equation}
Note that the operator ordering for this non-hermitian operator product is based on the convention used for the mathematical definition of complex weak values in Eq.(\ref{eq:wv}). In principle, it is also possible to represent only the real part of the weak values by using the self-adjoint part of the transient state, as defined by the average of the above operator ordering with the opposite operator ordering \cite{Hof10}. In the following, our main reason for using the operator representation for complex weak values is that its mathematical form is more simple and allows us to present the results in a more compact form. However, the same results could also be obtained with the self-adjoint transient states originally introduced in \cite{Hof10}. In the present context, it is sufficient that the transient states correctly describe the real parts of the weak values, and that the set of operators $\hat{T}(f)$ obtained for the different measurement outcomes $f$ can be recombined to obtain the original quantum state,
\begin{equation}
\label{eq:complete}
\hat{\rho}_i = \sum_f p(f) \hat{T}(f),
\end{equation}
where $p(f)$ is the probability of $f$ given by the product trace of $\hat{\rho}_i$ and $\hat{\Pi}_f$. The final measurement of $f$ thus decomposes the initial probability density $\hat{\rho}_i$ into sub-ensembles $\hat{T}(f)$ that represent the weak value statistics for each measurement outcome $f$. 

\begin{figure}[th]
\centering
\vspace{-4cm}
 \includegraphics[width=125mm]{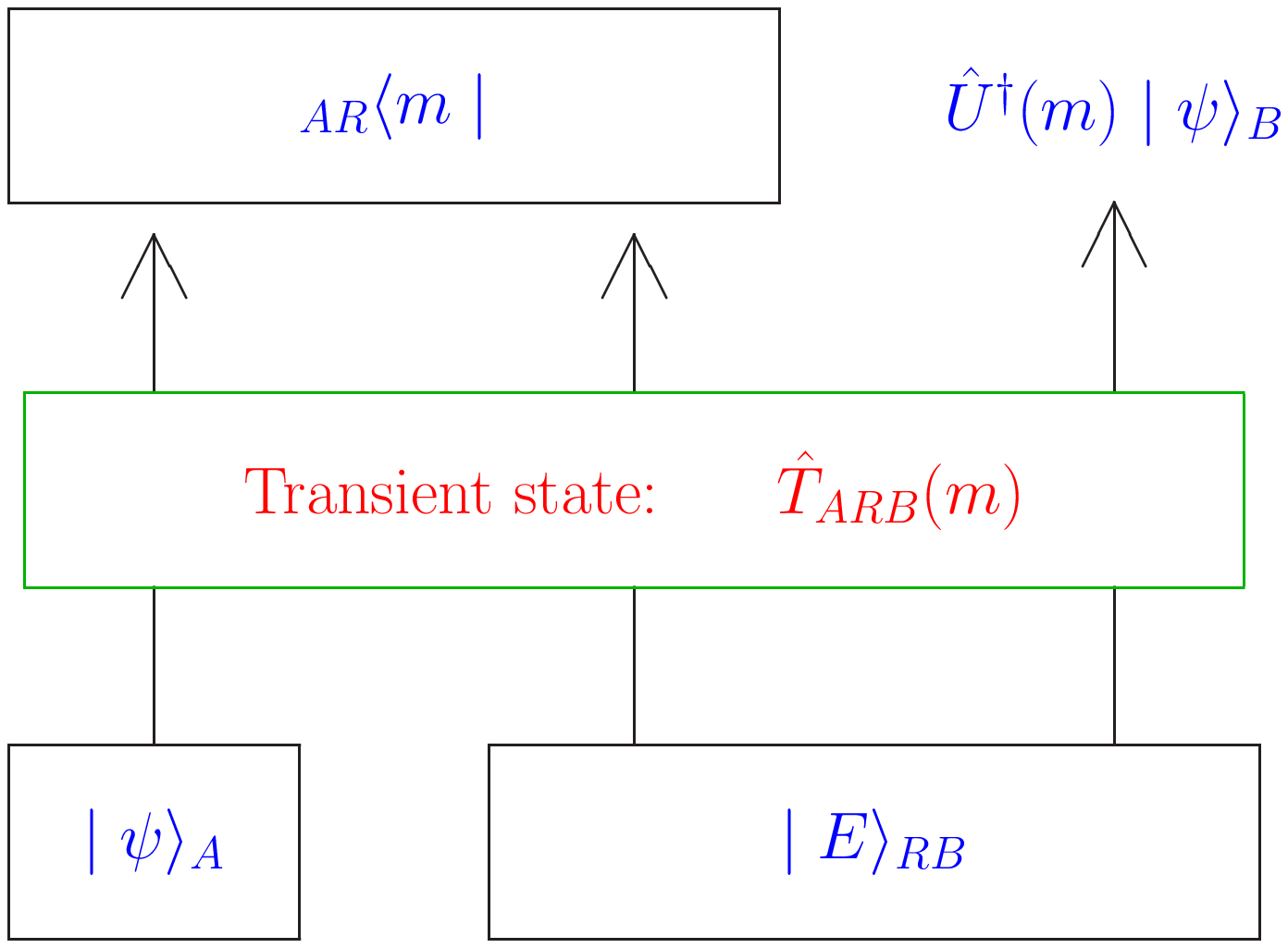}
\vspace{-6cm}
   \caption{Illustration of the transient state defined by a Bell measurement result of $m$. The state $\hat{T}_{ARB}(m)$ describes correlations between all three systems involved in the teleportation process.}
   \label{fig2}
\end{figure}

We can now apply the transient state analysis of weak measurement statistics to the case of quantum teleportation. The initial density matrix is given by the product of input state $\mid \psi \rangle \langle \psi \mid$ and entangled state $\mid E \rangle \langle E \mid$, and the Bell measurement is described by the product of a projection in $AR$ and the identity operation in $B$,
\begin{equation}
\hat{\Pi}(m) = \mid m \rangle \langle m \mid_{AR} \otimes \hat{I}_B.
\end{equation}
The weak measurement statistics of all three systems, $A$, $R$, and $B$, is then given by the transient state
\begin{equation}
\label{eq:TARB}
\hat{T}_{ARB}(m) = \frac{\mid \psi~;E\rangle\langle \psi~;E \mid \hat{\Pi}(m)}{\langle \psi~;E \mid \hat{\Pi}(m) \mid \psi~;E\rangle}
\end{equation}
As illustrated by Fig \ref{fig2}, the transient state $\hat{T}_{ARB}(m)$ describes the complete weak measurement statistics between the preparation of the entangled state and the Bell measurement. It is therefore possible to derive the outcomes of weak measurements for any observable $\hat{V}$ by simply taking the product trace of $\hat{V}$ and $\hat{T}_{ARB}(m)$. Importantly, the weak measurement statistics of the three systems do not factorize, indicating that the transient state describes correlations between the three systems. These correlations can be observed in weak measurements as the weak values of operators $\hat{V}$ constructed from products of local projection operators on the systems $A$, $R$, and $B$. In the following, we will take a closer look at these correlations and discuss the physics of teleportation in the light of these results. 

\section{Local weak measurement statistics}
\label{sec:local}

The compact formulation of the transient state $\hat{T}_{ARB}(m)$ given in Eq.(\ref{eq:TARB}) does not make use of the relation between the Bell state $\mid m \rangle$ and the initial entangled state defined in Eq.(\ref{eq:Bellstate}). Using this relation and the basis expansion of the initial entangled state in Eq.(\ref{eq:E}), the transient state can be expressed as
\begin{equation}
\label{eq:expandTARB}
\hat{T}_{ARB}(m) =\sum_{k,l} \mid \psi \rangle \langle k \mid \hat{U}^{\dagger}(m) \otimes
\mid l \rangle \langle k \mid \otimes
\mid l \rangle \langle \psi \mid \hat{U}(m).
\end{equation}
Note that the appearance of the unitary transform $\hat{U}(m)$ in the remote system $B$ is a consequence of the inner product of the projection on $\mid m \rangle_{AR}$ with the initial entangled state of $R$ and $B$, as shown in Eq.(\ref{eq:tele}). The transient operator thus explains the essential connection between the states in $A$ and $B$ in terms of a statistical correlation selected by the Bell measurement. These correlations can be observed in collective weak measurements of all three systems, where the weak measurements determine the weak values of an operator defined by products of local properties from all three systems. Simplifications are possible if the weak measurement only involves one of the three systems. In this case, the local transient states can be obtained from the partial traces over those systems that are not involved in the weak measurement. For local measurements of the input system $A$, the result reads 
\begin{equation}
\hat{T}_{A}(m) = \mbox{Tr}_{RB} \left(\hat{T}_{ARB}(m)\right) = \mid \psi \rangle \langle \psi \mid.
\end{equation}
Thus the weak values of the input system $A$ are equal to the expectation values of the input state, confirming that the post-selection of a Bell measurement outcome does not ``update'' the information about the input state in any way. Similarly, the state observed in local measurements of the output $B$ is given by
\begin{equation}
\hat{T}_{B}(m) = \mbox{Tr}_{AR} \left(\hat{T}_{ARB}(m)\right) = \hat{U}^\dagger(m) \mid \psi \rangle \langle \psi \mid \hat{U}(m).
\end{equation}
This is actually the correct output state of the teleportation, since the post-selection did not involve any measurement of system $B$, and weak values without post-selection of a final measurement result ($\hat{\Pi}_f = \hat{I}$) are necessarily equal to the expectation values observed in strong measurements. Significantly, local weak measurements on $A$ and on $B$ can be performed at the same time, before the post-selection of the Bell measurement. This means that the quantum state $\mid \psi \rangle$ appears to co-exist in both $A$ and $B$, suggesting a ``weak cloning'' effect as discussed in \cite{Sjo06}. 

It may also be interesting to look at the local state observed in the reference system $R$. Here, the trace over $A$ and $B$ has the curious effect of producing a complex conjugate of the initial state, given by
\begin{eqnarray}
\hat{T}_{R}(m) &=& \mbox{Tr}_{AB} \left(\hat{T}_{ARB}(m)\right)
\nonumber \\ &=& 
\left(\sum_{l} \langle \psi \mid \hat{U}(m) \mid l \rangle \mid l \rangle \right) \left(\sum_{k} \langle k \mid \langle k \mid \hat{U}^\dagger(m) \mid \psi \rangle \right).
\end{eqnarray}
Since complex conjugation in the energy basis is often associated with time reversal, it is possible to interpret $\hat{T}_{R}(m)$ as a signal running backward in time, as discussed in \cite{Llo11a,Llo11b}. In the context of ``weak cloning,'' however, the transient state of $R$ appears to describe the anti-clone usually associated with optimal cloning processes \cite{Sjo06,Fil04}.

\section{Correlations between input and output}
\label{sec:precorr}

The local weak measurement results show that the quantum information encoded in the coherence of the input state appears in all three systems equally. However, the transient state $\hat{T}_{ARB}(m)$ includes a more detailed description of the teleportation process in the form of correlations between the systems. To better understand the transfer of information between $A$ and $B$, we should consider the transient state 
\begin{equation}
\hat{T}_{AB}(m) = \sum_k  \mid \psi \rangle \langle k \mid \hat{U}^{\dagger}(m) \otimes
\mid k \rangle \langle \psi \mid \hat{U}(m),
\end{equation}
where only the reference system $R$ has been traced out. Clearly, the transient state $\hat{T}_{AB}(m)$ is different from the product of the local transient states, indicating that the sum over $k$ describes correlations between $A$ and $B$. 

To find out more about these correlations, we need to consider possible weak measurements of correlated observables. If we consider a local property $\hat{Q}$ with eigenstates $\mid q_i \rangle$, the weak measurement statistics can be expressed in terms of the weak values of the projection operators $\mid q_i \rangle\langle q_i \mid$. In strong measurements, these projection operators determine the probability of measuring the outcome $q_i$. The weak values of the projectors therefore correspond to the weak measurement equivalent of conditional probabilities \cite{Joh07,Hos10,Hof12a,Lun12,Tsu12}. Since the difference between the local transient states in $A$ and in $B$ is described by the unitary transform $\hat{U}(m)$, it is convenient to consider a measurement of $\hat{Q}$ in $A$ and a measurement of $\hat{U}^\dagger(m) \hat{Q} \hat{U}(m)$ in $B$. The weak values of the joint probabilities for the outcomes $\mid q_i \rangle$ in $A$ and $\hat{U}^\dagger(m) \mid q_j \rangle$ in $B$ is given by
\begin{equation}
\label{eq:idcorr}
\mbox{Tr}\left(\hat{T}_{AB}(m) \left(\mid q_i \rangle \langle q_i \mid \otimes \hat{U}^\dagger (m) \mid q_j \rangle \langle q_j \mid \hat{U}(m) \right)\right) = \delta_{i,j} \; |\langle q_i \mid \psi \rangle|^2.
\end{equation}
Although weak values of projection operators are generally complex and can have real values below zero or above one, the results obtained for the projectors of $\hat{Q}$ in $A$ and $\hat{U}^\dagger(m) \hat{Q} \hat{U}(m)$ in $B$ are all positive and correspond to a particularly simple classical probability distribution. Specifically, the Kronecker delta in Eq.(\ref{eq:idcorr}) indicates that the results in $A$ and in $B$ must always be equal, since the result is zero unless $i=j$. The probability of the outcome $i=j$ corresponds to the probability of the outcome $\mid q_i \rangle$ for the input state $\mid \psi \rangle$. Eq.(\ref{eq:idcorr}) therefore corresponds to the intuitive notion that system $A$ and system $B$ are identical copies of each other, where the state $\mid \psi \rangle$ describes the probability distribution of the perfectly correlated physical properties $\hat{Q}$.  

Importantly, the result of Eq.(\ref{eq:idcorr}) applies to all physical properties of the teleported system. Thus, the teleportation process can be interpreted as a faithful copy of all physical properties $\hat{Q}$, independent of input state. Note that this is very similar to the phase space picture obtained from the Wigner function representation of continuous variable teleportation, where the teleportation process faithfully transfers position and momentum, and the Wigner function merely describes a quantum statistical weight attached to each combination of position and momentum by the quantum state \cite{Bra98}. The role of the Bell measurement is the identification of the unitary $\hat{U}(m)$ that describes the deterministic relation between the physical properties in both systems. By selecting a complete set of $d^2$ Bell states, it is guaranteed that such a deterministic relation can be found in each measurement, indicating a close analogy between $d$-dimensional Hilbert spaces and discrete phase spaces described by a lattice of $d^2$ points \cite{Hof12a}. 

Effectively, weak measurements in $A$ and weak measurements (or strong measurements) in $B$ can be treated as measurements of the same system. Therefore, the correlations observed between different measurements in $A$ and in $B$ correspond to correlations of these properties within a single physical system. If we consider the measurement of two non-commuting observables $\hat{Q}$ and $\hat{P}$ with eigenstates $\mid q_i \rangle$ and $\mid p_j \rangle$, the complex joint probabilities defined by the weak values of the projectors is given by
\begin{eqnarray}
\label{eq:cjoint}
\lefteqn{\mbox{Tr}\left(\hat{T}_{AB}(m) \left(\mid q_i \rangle \langle q_i \mid \otimes \hat{U}^\dagger (m) \mid p_j \rangle \langle p_j \mid \hat{U}(m) \right)\right) =}
\nonumber \\[0.2cm] && \hspace*{7cm}
\langle p_j \mid q_i \rangle \langle q_i \mid \psi \rangle \langle \psi \mid p_j \rangle.
\end{eqnarray}
This result is identical to the joint probabilities reconstructed from weak measurements of $\hat{Q}$ followed by a post-selection of $\hat{P}$ \cite{Joh07,Hos10,Hof12a,Lun12}. Interestingly, the complex value of the weak measurement outcomes defines a specific measurement sequence. However, in the case of quantum teleportation this sequence does not correspond to the temporal order of measurements: clearly, it does not matter whether the measurement in $B$ is performed before or after the measurement in $A$. Instead, it is relevant that system $A$ is directly connected to the input, while $B$ is directly connected to the output. The connection between the two systems is not established by interactions, but by obtaining the necessary information about the accidental relation $\hat{U}(m)$ between the two systems. Since this information relates the past of system $B$ to the future of system $A$, it may seem as if the reference $R$ carries a signal backwards in time, as indicated by the transient state of the reference $R$. It is therefore possible to explain the appearance of ``closed time like curves'' discussed in \cite{Llo11a,Llo11b} in terms of the difference between the time at which information is obtained and the time to which this information applies. 

\section{Partial teleportation by quantum controlled Bell measurements}
\label{sec:partial}

According to the analysis presented above, quantum teleportation makes use of the accidental relation $\hat{U}(m)$ between the physical properties of $A$ and of $B$ to identify the transformation that makes the physical properties in the output of $B$ equal to the physical properties in the input of $A$. The Bell measurement merely identifies the relation between $A$ and $B$. This relation is valid for the complete time interval between entanglement generation and Bell measurement, so that system $B$ is a faithful copy of system $A$ during this time. It has been noted that this situation corresponds to a weak measurement version of cloning (``weak cloning''), where the usual limit of cloning fidelity does not apply and a cloning fidelity of one can be achieved \cite{Sjo06}. However, the Bell measurement completely randomizes the state of one of the clones, leaving behind only a single copy of the teleported state. 

An alternative approach to accessing the correlations between $A$ and $B$ in quantum teleportation is to reduce the measurement back-action of the Bell measurement to avoid the complete randomization of the input. However, this requires a reduction of the measurement resolution. Thus, there is a trade-off between the teleportation fidelity and the output fidelity of the original system after the Bell measurement, which ensures that the cloning fidelities remain within the established limits. It has already been shown that the partial teleportation implemented by a finite resolution Bell measurement with minimal back-action results in an optimal cloning scheme \cite{Fil04}. Here, we generalize the discussion by considering coherent superpositions of the Bell measurement and the identity operation. We can then show that the transient state $\hat{T}_{ARB}$ is part of the actual output state of a partial teleportation and contributes to the optimal cloning fidelity for the outputs of $A$ and $B$. 

A partial Bell measurement can be defined as a quantum coherent superposition of a fully projective measurement of $\mid m \rangle$ and a random assignment of the measurement outcome that leaves the state unchanged. The measurement operator acting on the initial state can then be written as
\begin{equation}
\hat{M}(m) = C_M \mid m \rangle\langle m \mid + C_I \frac{1}{d} \hat{I}\otimes \hat{I},
\end{equation}
where $C_M$ and $C_I$ are the coherent amplitudes associated with measurement and identity operation, respectively. Since the positive operator valued measure given by $\hat{M}^\dagger(m) \hat{M}(m)$ should be normalized to give a total probability of one for all possible measurement outcomes, the coherent amplitudes $C_M$ and $C_I$ can be determined by a single parameter, e.g.
\begin{eqnarray}
C_M (\theta) &=& \frac{\cos\theta}{\sqrt{1+\frac{2}{d}\sin\theta \cos\theta}},
\nonumber \\
C_I (\theta) &=& \frac{\sin \theta}{\sqrt{1+\frac{2}{d}\sin\theta \cos\theta}}.
\end{eqnarray}
Here, the parameter $\theta$ determines the reduction of measurement strength, with $\theta=0$ indicating a fully projective measurement and $\theta=\pi/2$ indicating a random assignment of measurement outcome that does not change the quantum state. 

\begin{figure}[th]
 \centering 
\vspace{-3cm}
\includegraphics[width=125mm]{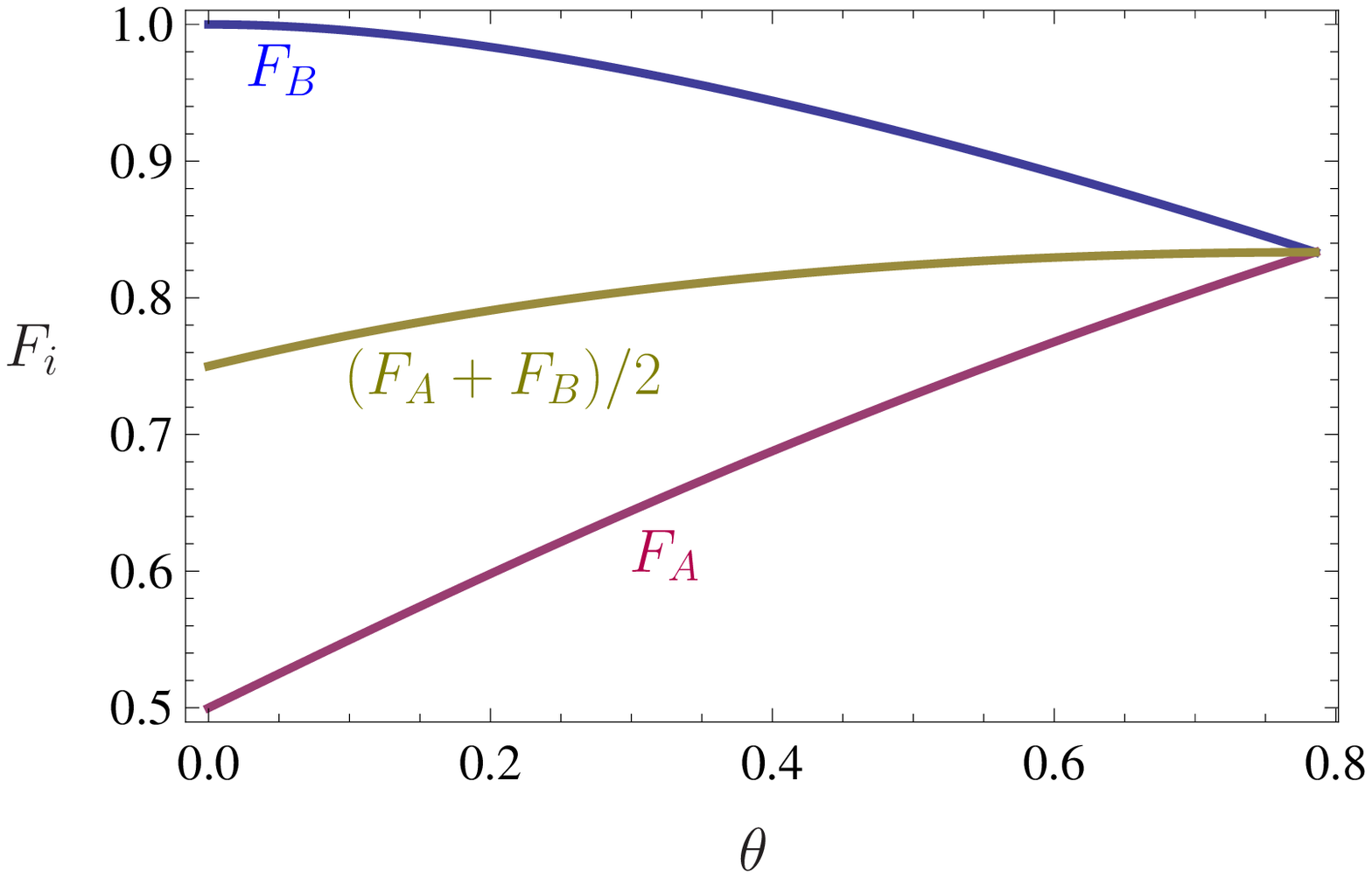}
\vspace{-6.5cm}
 \caption{Transition between optimal teleportation and optimal cloning as a function of the reduction of Bell measurement strength described by the parameter $\theta$. As Bell measurement strength is reduced, the fidelity $F_A$ of the input system $A$ after the Bell measurement increases and the teleportation fidelity $F_B$ decreases until the two values become equal at $\theta=\pi/4$. The average fidelity reaches its maximal possible value at this point, resulting in an implementation of optimal cloning as discussed in \cite{Fil04}.}
\label{fig3}
\vspace{0.5cm}
\end{figure}

The effects of the partial Bell measurement on the input state $\mid \psi; E \rangle$ are found by applying the operators $\hat{M}(m)$ to systems $A$ and $R$ of the initial state,
\begin{eqnarray}
\lefteqn{\left(\hat{M}_{AR}(m) \otimes \hat{I}_B \right) \mid \psi; E \rangle_{ARB}
=} \nonumber \\ \hspace*{2cm}
C_M \hat{\Pi}(m) \mid \psi; E \rangle_{AR;B} + C_I \frac{1}{d} \mid \psi; E \rangle_{ARB}.
\end{eqnarray}
Thus, the output state is a quantum coherent superposition of the initial state and the final state of the ideal teleportation process. It is possible to write the density matrix of this output state as an effective mixture of three components, corresponding to the density matrix of the initial state, the density of the final state, and the self-adjoint part of the transient state,
\begin{eqnarray}
\hat{\rho}^{\mathrm{out}}_{ARB}(m) &=& \rho_i \mid \psi; E \rangle \langle \psi; E \mid + 
\rho_f d^2 \hat{\Pi}(m) \mid \psi; E \rangle \langle \psi; E \mid  \hat{\Pi}(m) 
\nonumber \\ &&
+ \rho_T \frac{1}{2}\left(\hat{T}_{ARB}(m) + \hat{T}_{ARB}^\dagger(m)\right).
\end{eqnarray}
The statistical weights of the three contributions are given by 
\begin{eqnarray}
\rho_i &=& \hspace{0.3cm} C_I^2,
\nonumber \\
\rho_f &=& \hspace{0.3cm} C_M^2,
\nonumber \\
\rho_T &=& \frac{2}{d} C_I C_M.
\end{eqnarray}
The quantum statistics of the output can therefore be explained in terms of the statistics of initial state, final state, and transient state. In particular, the teleportation fidelity $F_B$ (obtained after the application of $\hat{U}(m)$) is equal to one in the final state and in the transient state, while it is $1/d$ in the initial state. Likewise, the fidelity $F_A$ of the output state of system $A$ after the partial Bell measurement is one for the initial state and for the transient state, but only $1/d$ for the final state. The fidelities can therefore be given in terms of the statistical weights of the three contributions to the output state,
\begin{eqnarray}
\label{eq:fid}
\hspace*{0.5cm} F_B &=& 1-\frac{d-1}{d} \rho_i, 
\nonumber \\
\hspace*{0.5cm} F_A &=& 1-\frac{d-1}{d} \rho_f,
\nonumber \\
\frac{1}{2}(F_A+F_B) &=& \frac{d+1}{2d} + \frac{d-1}{2d} \rho_T.
\end{eqnarray}
As discussed in \cite{Fil04}, partial teleportation is an optimal cloning procedure. In particular, the symmetric case of $\theta=\pi/4$ corresponds to optimal $1 \to 2$ cloning. Fig. \ref{fig3} illustrates the transition between optimal teleportation at $\theta=0$ and optimal cloning at $\theta=\pi/4$ for the case of a two level system ($d=2$). As the strength of the Bell measurement is reduced, the teleportation fidelity drops and the fidelity of the input system after the Bell measurement increases. 

For the present discussion, the most important aspect of the partial teleportation is that it permits a direct measurement of correlations between $A$ and $B$. In fact, it has already been shown in previous work that optimal cloning of the type realized at $\theta=\pi/4$ includes an ideal cloning term that reproduces the joint probabilities of Eq.(\ref{eq:cjoint}) as predicted from weak measurements \cite{Hof12b}. In the general case of partial teleportation, the joint probabilities obtained in measurements of $\hat{Q}$ in the output of $A$ and $\hat{P}$ in the output of $B$ are given by 
\begin{eqnarray}
p(q_i,p_j) &=& \frac{1}{d} |\langle q_i \mid \psi \rangle|^2 \rho_i 
+ \frac{1}{d} |\langle p_j \mid \psi \rangle|^2 \rho_f 
\nonumber \\[0.1cm] &&
+ 
\mbox{Re}\left(\langle p_j \mid q_i \rangle \langle q_i \mid \psi \rangle \langle \psi \mid p_j \rangle \right) \rho_T.
\end{eqnarray}
Thus, a partial Bell measurement preserves a non-vanishing contribution from the real part of the weak measurement statistics that can be observed before the Bell measurement is performed. Indeed, the complete weak measurement statistics can even be obtained in the limit of $\theta \to 0$, where $\rho_i$ is negligibly small. In this case, the weak value given in Eq.(\ref{eq:cjoint}) can be obtained from the small differences between the joint probabilities $p(q_i,p_j)$ for different values of $q_i$. The logic of this measurement is surprisingly similar to that of the weak measurement, with the essential difference that the low resolution of the measurement originates from the nearly maximal disturbance of the state by the Bell measurement, and not from the weakness of the measurement interaction. 

\section{Physics of quantum teleportation}
\label{sec:discuss}

The picture of quantum teleportation that has emerged from the weak measurement analysis and its application to partial teleportation indicates that the physical properties of the input system $A$ and the remote system $B$ are related by the unitary $\hat{U}(m)$ even before the Bell measurement has taken place. This is possible because the initial state already includes all possible relations $\hat{U}(m)$ as a coherent superposition. In the weak measurement analysis, this superposition can be expressed as an effective mixture of the $d^2$ possibilities according to Eq.(\ref{eq:complete}). The Bell measurement merely selects the possibility valid in each specific teleportation, without changing the physical properties of the remote system at all. 

Weak measurements indicate that the physical properties of system $B$ originate from the source of the initial entanglement and experience no change or modification during the Bell measurement. Since the physical properties of the input state do not change until they are modified by the measurement back-action of the Bell measurement, the input properties of $A$ and the output properties of $B$ appear to coexist in the interval between entangled state preparation and Bell measurement. The relation $\hat{U}(m)$ that relates the physical properties of $B$ to the physical properties of $A$ is initially random. In fact, the randomness of the local state of $B$ in the entangled state guarantees that each relation $m$ is equally likely, regardless of the input state in $A$. Teleportation is not achieved by a physical interaction, but by obtaining information about this random relation between $A$ and $B$ in the Bell measurement. Specifically, the combination of the pre-determined correlations in $\mid E \rangle$ with the correlations in $\mid m \rangle$ discovered in the Bell measurement results in the knowledge that $\hat{U}(m)$ describes the relation between the unmodified physical properties in $A$ and in $B$. 

The only reason why the perfect correlations between $A$ and $B$ can never be observed directly is the measurement back-action of the Bell measurement, which completely randomizes the local properties of system $A$. However, a partial Bell measurement can be used to avoid complete randomization, and the correlations between $A$ and $B$ observed after such a partial teleportation correspond to the ones obtained in the weak measurements performed before a full Bell measurement. Our analysis thus illustrates the conceptual relation between the ``weak cloning'' observed in teleportation and the optimal cloning realized by partial teleportation \cite{Sjo06,Fil04}. In both cases, the precise relation between the physical properties of the input $A$ and the output $B$ is determined by the correct result of the Bell measurement. In the ``weak cloning'' limit, the Bell measurement is precise, but the state of $A$ is randomized after the measurement. In the partial teleportation limit, there is still some randomization in $A$ and the reduced measurement resolution results in teleportation errors in $B$, but the error free contribution that remains is precisely described by the self-adjoint part of the transient state observed in weak measurements.

\section{Conclusions}
\label{sec:conclude}
Weak measurement statistics provide a powerful tool for the analysis of quantum processes. In the case of teleportation, it is possible to discover close analogies between classical statistics and quantum statistics that are difficult to recognize when the conventional measurement postulate is applied. In particular, weak measurement statistics can separate the information about the past obtained in the Bell measurement from the back-action effects associated with the measurement. It is then possible to show that the change of the quantum state in $B$ originates from the selection of a sub-ensemble that was already present before the measurement. Thus, quantum teleportation does not involve any ``spooky'' action at a distance. The Bell measurement at $A$ merely identifies the correct transform $\hat{U}(m)$ that relates the physical properties of $B$ to the physical properties $A$ even before the Bell measurement was performed. 

Significantly, the relation $\hat{U}(m)$ can be applied to all physical properties, and the correlations described by the weak values of projection operators in $A$ and in $B$ suggest that all properties of $A$ are perfectly correlated with the corresponding properties in $B$. Teleportation thus copies all physical properties of the quantum system $A$ equally, irrespective of whether the input state is an eigenstate of the property or not. Effectively, the relation defined by $\hat{U}(m)$ is more fundamental to teleportation than the specific values taken by the physical properties themselves. By using weak measurement statistics, it is therefore possible to verify that the quantum fluctuations of physical properties $\hat{Q}$ in $A$ and $B$ are perfectly correlated without having to assign counterfactual values to $\hat{Q}$ in either system. Teleportation can thus be understood as an operation on the physical properties of the systems, where the Bell measurement obtains the relevant information about the relation $\hat{U}(m)$ that maps the properties of $A$ onto the properties of $B$.

\section*{Acknowledgments}
This work was supported by JSPS KAKENHI Grant Number 24540427.

\end{document}